\documentclass[11pt]{article}
\newif\iffigs\figstrue

\usepackage[title]{appendix}
\usepackage{epsfig,latexsym}
\usepackage{amsmath}
\usepackage{verbatim}
\usepackage{color}
\usepackage{mathrsfs}
\usepackage{amssymb}

\DeclareMathAlphabet{\mathpzc}{OT1}{pzc}{m}{it}

 \csname
@addtoreset\endcsname{equation}{section}

\def\gz0{\gamma^{0}}

\def\beq{\begin{equation}}
\def\eeq{\end{equation}}
\def\bea{\begin{eqnarray}}
\def\eea{\end{eqnarray}}
\def\ba{\begin{array}}
\def\ea{\end{array}}
\def\bec{\begin{center}}
\def\ec{\end{center}}
\def\ba{\begin{align}}
\def\ena{\end{align}}

\def\12{\frac{1}{2}}

\begin{document}

\begin{flushright}
{\today}
\end{flushright}

\vspace{10pt}

\begin{center}


{\Large\sc On Boundaries, Charges and Fermi Fields}\\


\vspace{25pt}
{\sc J.~Mourad${}^{\; a}$  \ and \ A.~Sagnotti${}^{\; b}$\\[15pt]

${}^a$\sl\small APC, UMR 7164-CNRS, Universit\'e de Paris \\
10 rue Alice Domon et L\'eonie Duquet \\75205 Paris Cedex 13 \ FRANCE
\\ e-mail: {\small \it
mourad@apc.univ-paris7.fr}\vspace{10
pt}

{${}^b$\sl\small
Scuola Normale Superiore and INFN\\
Piazza dei Cavalieri 7\\ 56126 Pisa \ ITALY \\
e-mail: {\small \it sagnotti@sns.it}}\vspace{10pt}
}

\vspace{40pt} {\sc\large Abstract}\end{center}
\noindent We address some general issues related to torsion and Noether currents
for Fermi fields in the presence of boundaries, with emphasis on the
conditions that guarantee charge conservation.
We also describe exact solutions of these boundary conditions and some implications for string vacua with broken supersymmetry.

\setcounter{page}{1}

\pagebreak

\newpage
\newpage
\baselineskip=20pt
\section{\sc  Introduction}\label{sec:intro}

String compactifications have been widely explored during the last decades, but almost exclusively with closed internal manifolds~\cite{stringtheory}, so that the boundary conditions needed for Fermi fields when the manifold has a border have received little attention. Two notable exceptions are the Neveu--Schwarz--Ramond (NSR) open string~\cite{NSR} and the Horava--Witten link~\cite{HW} between the $E_8 \times E_8$ heterotic string and the Cremmer--Julia--Scherk~\cite{CJS} eleven--dimensional form of Supergravity~\cite{SUGRA}. Boundaries, however, have played so far a prominent role in vacuum configurations for orientifolds~\cite{orientifolds} with ``brane supersymmetry breaking''~\cite{bsb,dmnonlinear}, whose prototype is the nine--dimensional Dudas--Mourad solution of~\cite{dm_vacuum}. This involves regions of strong coupling, but is classically stable~\cite{bms} and the tension from branes and orientifolds, which signals the breaking of supersymmetry, renders the length of its internal interval finite. This compactification also concerns the $U(32)$ non--supersymmetric orientifold of~\cite{AS95}, while a variant~\cite{dm_vacuum} applies to the non--supersymmetric heterotic model of~\cite{AGGMV}. These examples motivate, in our view, a closer look at their Fermi fields.

For definiteness, we choose a coordinate system such that the boundary $\partial {\cal M}$ of the $D$--dimensional manifold ${\cal M}$ lies at $r=0$ and the metric takes nearby the form
\beq
ds^2 \ \equiv \ g_{MN}
\, dx^M\,dx^N \ =  \ g_{rr}\, dr^2 \ + \ ds^2_\perp \ . \label{boundary_metric}
\eeq
The variation of the Dirac action for a spinor $\lambda$ yields boundary terms, which can be removed provided~\footnote{We use a ``mostly plus'' signature, so that $\gamma^0$ is antihermitian while the other $\gamma$--matrices are hermitian.}
\beq
\left.{}^{} \left(\bar\lambda \gamma^r\delta\lambda-\delta\bar\lambda\gamma^r\lambda\right)\right|_{\partial\,{\cal M}} \ = \ 0 \ . \label{boundary_term}
\eeq
Any boundary condition
\beq
\left. {}^{}\left(1 \ - \ \Lambda\right) \lambda\right|_{\partial{\cal M}} \ = \ 0 \ , \label{Lambda_mat}
\eeq
with a Hermitian matrix $\Lambda$ such that
\beq
\Lambda^2 \ = \ 1\ , \qquad \left\{ \Lambda \,,\, \gamma^0\gamma^r\right\} \ = \ 0 \label{Lambda_constr_kin}
\eeq
solves eq.~\eqref{boundary_term}.
Different choices are possible, however, depending on the symmetries to be preserved:
for example, $\Lambda=\gamma^r$ and $\Lambda=i\gamma^0$ are two solutions, and there are more options. One of our aims is to connect the allowed choices of $\Lambda$ to the conservation of Noether Killing charges.

In Sections~\ref{sec:bianchi_b} and \ref{sec:bianchi_f} we discuss the matter and gravity Bianchi identities related to diffeomorphisms and local Lorentz symmetries, taking into account that the back--reaction of Fermi fields includes in general the emergence of torsion. In Section~\ref{sec:currents} we connect diffeomorphisms and local Lorentz Bianchi identities to Noether Killing currents for global isometries, whose normal components should vanish on the boundary $\partial{\cal M}$ to grant charge conservation. This places further constraints on $\Lambda$, which we explore in Section~\ref{sec:applications} with an eye to string models with broken supersymmetry.

\vskip 12pt

\section{\sc  Bianchi Identities and Bose Fields}\label{sec:bianchi_b}

Let us begin by reviewing briefly the behavior of Bose fields with reference to the simplest case, a real scalar $\phi$. If the metric takes the form~\eqref{boundary_metric} near the boundary $\partial{\cal M}$ of a $D$--dimensional manifold ${\cal M}$, the variation of the standard kinetic term yields the boundary condition
\beq
\left. \delta \phi \ \partial_r \phi \right|_{\partial\,{\cal M}} \ = \ 0  \label{neu_dir} \ ,
\eeq
which is solved by the familiar Neumann $(\partial_r \phi=0)$ or Dirichlet $(\delta\phi=0)$ choices. Notice that the latter only implies that $\phi$ is a fixed function on $\partial{\cal M}$. Similar remarks apply to forms and to the metric tensor, up to Gibbons--Hawking terms~\cite{gh}.

Let us now explore whether eq.~\eqref{neu_dir} suffices to guarantee the conservation of Noether Killing charges, which are built from \emph{symmetric} energy--momentum tensors $T^{MN}$ defined via the metric variations
\beq
\delta{\cal S}_m \ = \ \int_{\cal M} d^{D}x \, \sqrt{-g}\ \delta g_{MN}\ T^{MN} \ . \label{emt_bose}
\eeq
A consistent coupling to gravity demands that $\delta{\cal S}$ vanish for the metric variations
\beq
\delta\,g_{MN} \ = \ D_M\,\xi_N \ + \ D_N\,\xi_M  \ , \label{delta_g}
\eeq
which describe the effect of diffeomorphisms $\delta x^M = \xi^M$ when keeping fixed the coordinates in fields, and with $\xi^M$ of local support a partial integration leads to the Bianchi identity
\beq
D_M\, T^{MN} \ = \ 0 \ \label{cons_bose} .
\eeq
Continuous symmetries of $g_{MN}$ are generated by Killing vectors $\zeta^M$, solutions of~\eqref{delta_g} with $\delta g_{MN}=0$, and lead to the covariantly conserved Noether currents
\beq
{\cal J}^M \ = \ T^{MN}\,\zeta_N  \label{J_bose} \ .
\eeq
The combinations $\sqrt{-g}\, {\cal J}^M$ satisfy the ordinary conservations law $\partial_M \left(\sqrt{-g}\, {\cal J}^M\right)=0$, and in the absence of a boundary the charges $Q(t)$, which we write for brevity in the form
\beq
Q(t) \ = \ \int_{\cal M}  d^{D} x \, \delta(x^0 - t)\,\sqrt{-g} \ {\cal J}^0 \ , \label{charges}
\eeq
are conserved. However, when ${\cal M}$ has a boundary $\partial {\cal M}$
\beq
\frac{d\,Q(t)}{dt} \ = \ \int_{\partial{\cal M}}  d^{D-1} x \, \delta(x^0 - t)\,\sqrt{-g} \ {\cal J}^r \ ,
\eeq
and the condition
\beq
\left. {\cal J}^r \right|_{\partial {\cal M}} \ \equiv \ \left. T^{\,rN}\,\zeta_N \right|_{\partial {\cal M}} \ = \ 0 \ , \label{bc_bose}
\eeq
is needed to prevent charge flow across the boundary. It involves off--diagonal components of the energy--momentum tensor since $\zeta^r$ {should vanish} on $\partial{\cal M}$ in order not to affect it.
For the bosonic actions of interest, the boundary conditions like~\eqref{neu_dir} that emerge from the equations of motion must be supplemented in general by eq.~\eqref{bc_bose}. For instance, Killing translation symmetries on $\partial{\cal M}$ require for a Dirichlet scalar $\phi$ that
\beq
\left. \zeta^M\,\partial_M \phi \right|_{\partial{\cal M}} \ = \ 0\ ,
\eeq
whereas for a Neumann scalar eq.~\eqref{bc_bose} is identically satisfied.

\section{\sc Bianchi Identities and Fermi Fields}\label{sec:bianchi_f}

When Fermi fields are present, local Lorentz transformations also acquire a key role, and there are consequently a few novelties. The metric tensor leaves way to the vielbein ${e_M}^A$ and the spin connection ${\omega_M}^{AB}$, while the variation of the matter action,
\beq
\delta\,{\cal S}_m \ = \ \int_{\cal M} d^{D}x \ e \ \left[ \delta {e_M}^A\, {{\cal T}^{M}}_A \ + \ \delta {\omega_M}^{AB} \ {{\cal Y}^M}_{AB} \right] \ , \label{emt_fermi}
\eeq
now defines generally a \emph{non--symmetric} energy--momentum tensor ${{\cal T}^{M}}_A$ and a new tensor ${{\cal Y}^M}_{AB}$. In the following, early Latin labels describe flat indices, while late Latin labels describe curved ones. The vielbein is covariantly constant,
\beq
D_M\,{e_N}^A \ \equiv \ \partial_M\,{e_N}^A \ + \ {{\omega}_M}^{AB}\ e_{NB} \ - \ {\Gamma^P}_{MN} \, {e_P}^A \ = \ 0 \ ,
\eeq
and this condition defines the ${\Gamma^P}_{MN}$, whose antisymmetric part
\beq
{S^{P}}_{MN} \ = \ {\Gamma^P}_{MN} \ - \ {\Gamma^P}_{NM}
\eeq
is the torsion tensor.

A local Lorentz transformation with parameters $\epsilon^{AB}=-\,\epsilon^{BA}$ acts as
\beq
\delta\, {e_M}^A \ = \ \epsilon^{AB}\, {e_{MB}} \ , \qquad \delta\, {\omega_M}^{AB} \ = \ - \ D_M\,\epsilon^{AB} \ .
\eeq
Rephrasing the argument reviewed for Bose fields, eq.~\eqref{emt_fermi} yields
\beq
\delta\,{\cal S}_m \ = \ \int d^{D}x \, e \ \left[ \epsilon^{AB}\, {e_{MB}}\,{{\cal T}^{M}}_A \ - \ D_M\,\epsilon^{AB} \ {{\cal Y}^M}_{AB} \right] \ ,
\eeq
and after a partial integration one obtains the Bianchi identity
\beq
D_M\, {{\cal Y}^M}_{AB} \ - \ {S^P}_{PM} \,{{\cal Y}^M}_{AB} \ = \ \frac{1}{2} \left( {{\cal T}}_{AB} \ - \ {{\cal T}}_{BA} \right) \ . \label{bianchi1}
\eeq
This step entails a small subtlety, since in the presence of torsion the covariant derivative of a vector $V^M$, equal to $\epsilon^{AB}{{\cal Y}^M}_{AB}$ in this case, does not lead to a total derivative, but
\beq
D_M\,V^M \ = \ {S^{M}}_{MN}\, V^N \ + \ \frac{1}{e}\ \partial_M\left( e\, V^M \right)  \ . \label{tot_der}
\eeq

Up to a local Lorentz rotation, diffeomorphisms act on ${e_M}^A$ and ${\omega_M}^{AB}$ as
\beq
\delta\,{e_M}^A \ = \ D_M\, \xi^A \ - \ {S^A}_{MN} \, \xi^N \ , \qquad \delta\, {\omega_M}^{AB} \ = \ - \ {R_{MN}}^{AB} \, \xi^N \ ,
\eeq
when keeping fixed the coordinates in fields, where we define the Riemann tensor, following the conventions in~\cite{Wald}, as
\bea
{R_{MN}}^{AB} &=& \partial_M\, {\omega_N}^{AB} \ - \ \partial_N\, {\omega_M}^{AB} \ + \ {\omega_M}^{AC}\, {{\omega_N}_C}^{B}\ - \ {\omega_N}^{AC}\, {{\omega_M}_C}^{B} \nonumber \\
&=& e^{PB}\,{e}^{QA}\, \left( \partial_N\,{\Gamma^P}_{MQ} \ - \ \partial_M\,{\Gamma^P}_{NQ}  \ + \ {\Gamma^P}_{NR}\, {\Gamma^R}_{MQ} \ - \ {\Gamma^P}_{MR}\, {\Gamma^R}_{NQ} \right) \ .
\eea
Resorting again to~\eqref{tot_der}, a partial integration now leads to a second Bianchi identity,
\beq
D_M\,{{\cal T}^M}_N \ + \ {S^P}_{MN}\,{{\cal T}^M}_P - \ {S^P}_{PM} \,{{\cal T}^M}_{N}  \ = \ - \ {R_{MN}}^{AB}\ {{\cal Y}^M}_{AB} \ . \label{bianchi2}
\eeq

For a spin--$\frac{1}{2}$ Fermi field $\lambda$ the Hermitian Dirac action
\beq
{\cal S}_m \ = \ - \ \frac{i}{2} \, \int d^{D} x \ e \ \left[ \bar{\lambda}\ \gamma^M
\, D_M \, \lambda \ - \ D_M\,\bar{\lambda}\, \gamma^M \, \lambda \right]
\eeq
determines
\beq
{{\cal T}^M}_A \ = \ \frac{i}{2} \left[ \bar{\lambda}\, \gamma^M\, D_A\,\lambda \ - \ D_A\,\bar{\lambda}\,\gamma^M \, \lambda \right] \ , \qquad
{{\cal Y}^M}_{AB} \ = \ - \ \frac{i}{4} \ \bar{\lambda} \, \gamma_{ABC}\,\lambda  \ e^{MC} \ , \label{TY_onehalf}
\eeq
where we have kept in ${\cal T}$ only terms that do not vanish on shell.
The boundary condition is now eq.~\eqref{boundary_term}, and in this case ${\cal Y}$ is totally antisymmetric, so that the traces ${S^M}_{MA}$ are absent in eqs.~\eqref{bianchi1}, \eqref{tot_der} and \eqref{bianchi2}.
However, they play a role for a spin--$\frac{3}{2}$ Fermi field $\psi_M$, since the Hermitian Rarita--Schwinger action
\beq
{\cal S}_m \ = \ - \ \frac{i}{2} \, \int d^{D} x \ e \ \left[ \bar{\psi}_M\ \gamma^{MNP}
\, D_N \, \psi_P \ - \ D_N\,\bar{\psi}_M\, \gamma^{MNP} \, \psi_P \right]
\eeq
determines
\bea
{{\cal T}^M}_A &=& \frac{i}{2} \left[ D_A\,\bar{\psi}_N\, \gamma^{MNP} \, \psi_P \ - \ \bar{\psi}_N\ \gamma^{MNP}
\, D_A \, \psi_P \ \right] \ , \nonumber \\
{\cal Y}^{MAB} &=& \frac{i}{4} \ \bar{\psi}_N \, \gamma^{MNPAB} \,\psi_P \ - \ \frac{i}{4}\left[ \bar{\psi}^A\,\gamma^M\,\psi^B \ + \  \bar{\psi}_N\,\gamma^N\,\psi^A \ e^{MB} \right. \nonumber \\  &+& \bar{\psi}^A\,\gamma^P\,\psi_P \ e^{MB} \ - \  \left. \left( A \leftrightarrow B\right)
\right] \ , \label{gravitino_boundaries}
\eea
and consequently
\beq
{{{\cal Y}^{M}}_M}{}^B \ = \ i\,\frac{(D-2)}{4} \left( \bar{\psi}_M\,\gamma^M\,\psi^B \ - \ \bar{\psi}^B\,\gamma^M\,\psi_M \right) \ .
\eeq
In ${\cal T}$ we have kept again only terms that do not vanish on shell, and
the counterpart of the boundary conditions~\eqref{boundary_term} and \eqref{neu_dir} is now
\beq
\left. \left( \bar{\psi}_M\ \gamma^{MrP}
\, \delta\,\psi_P \ - \ \delta\, \bar{\psi}_M\ \gamma^{MrP}
\, \psi_P \right) \right|_{\partial{\cal M}}  \ = \ 0 \ . \label{bc_threehalfs}
\eeq

In a similar fashion, varying the vielbein and the spin connection in the Einstein--Hilbert action
\beq
{\cal S}_{EH} \ = \ \frac{1}{2\,k^2}\ \int_{\cal M} d^{D}x \ e \ {e^M}_A\,{e^N}_B\,{R_{MN}}^{AB} \eeq
yields
\beq
\delta\,{\cal S}_{EH}  \ = \  - \ \frac{1}{k^2} \int_{\cal M} d^{D}x \ e \left[ \delta\,{\omega_N}^{AB} \,{\Theta^N}_{AB} \ + \ \delta\,{e_M}^A\,{G^M}_A \right] \ ,
\eeq
where
\bea
{G^M}_A  \,=\, \left({e^M}_C\,{e^P}_A \, - \, \frac{1}{2}\,{e^M}_A\,{e^P}_C \right) {e^Q}_D\, {R^{CD}}_{PQ}\,=\, {R^M}_A \, - \, \frac{1}{2}\, {e^M}_A\, R
\eea
is generally a \emph{non--symmetric} Einstein tensor, and
\beq
{\Theta^N}_{AB} \ = \ - \ \frac{1}{2}\left({S^P}_{PA} \ {e^N}_B \ - \ {S^P}_{PB} \ {e^N}_A \right) \ - \ \frac{1}{2}\ {S^N}_{AB} \ . \label{theta_s}
\eeq
Retracing the preceding arguments leads to the Bianchi identities
\bea
&& D_M\, {\Theta^M}_{AB} \ - \ {S^P}_{PM} \,{\Theta^M}_{AB} \ = \ \frac{1}{2} \left( {G}_{AB} \ - \ {G}_{BA} \right)  \ , \nonumber \\
&& D_M\,{G^M}_N \ + \ {S^P}_{MN}\,{G^M}_P\ - \ {S^P}_{PM} \,{G^M}_{N}  \ = \ - \ {\Theta^M}_{AB}\,{R_{MN}}^{AB} \ , \label{bianchi_grav}
\eea
that reflect the invariance of the Einstein--Hilbert Lagrangian under local Lorentz transformations and diffeomorphisms, while putting together matter and gravity sectors leads to the equations of motion
\beq
{G^M}_A \ = \ 2\, k^2\, {{\cal T}^M}_A \ , \qquad
{\Theta^M}_{AB} \ = \ 2\, k^2\,{{\cal Y}^M}_{AB} \ ,
\eeq
which are manifestly compatible with the Bianchi identities of eqs.~\eqref{bianchi1}, \eqref{bianchi2} and \eqref{bianchi_grav}.
Notice, finally, that eqs.~\eqref{bianchi_grav} would follow directly from the Bianchi identities for the Riemann tensor,
\bea
&& {R_{[MNP]}}^A \ = \ D_{[M}\,{S^A}_{NP]} \ - \ {S^R}_{[MN}\,{S^A}_{P]R} \ , \nonumber \\
&& D_{[M}\,{R_{NP]}}^{AB} \ =  \ {S^R}_{[MN}\, {R_{P]R}}^{AB}  \ ,
\eea
here expressed in terms of covariant derivatives including the torsion contribution, under which the vielbein is covariantly constant.

\section{\sc  Killing Vectors and Fermi Fields}\label{sec:currents}

\vskip 12pt

In the presence of Fermi fields, continuous symmetries and Killing vectors are to be defined with reference to diffeomorphisms, with parameters $\zeta^M$, and local Lorentz rotations, with parameters $\theta^{AB}$, whose combined effects leave both $e$ and $\omega$ invariant. These two conditions read
\bea
\delta\,{e_M}^A &\equiv& \ D_M\, \zeta^A \ - \ {S^A}_{MN} \, \zeta^N \ + \ \theta^{AB}\, e_{NB} \ = \ 0 \ , \nonumber \\
\delta\, {\omega_M}^{AB} &\equiv& - \ {R_{MN}}^{AB} \, \zeta^N \ - \ D_M\,\theta^{AB} \ = \ 0 \ , \label{deltaeomega}
\eea
and the first determines
\beq
\theta^{AB} \ = \ D^A\, \zeta^B \ - \ {S^{BA}}_C\, \zeta^C \ , \label{epsilonAB}
\eeq
while the antisymmetry of $\theta^{AB}$ translates into the \emph{modified Killing equation}
\beq
D_M\,\zeta_N \ + \ D_N\,\zeta_M \ = \ \left( {S_{MN}}^P \ + \ {S_{NM}}^P\right)\zeta_P  \ . \label{mod_killing}
\eeq
Moreover, using eq.~\eqref{epsilonAB}, the second of eqs.~\eqref{deltaeomega} can be cast in the form
\beq
D_M\,D_A\,\zeta_B \ = \ \left(D_M\,{S_{BA}}^N\right) \zeta_N \ + \ {S_{BA}}^N\,D_M\,\zeta_N \ - \ R_{MNAB}\,\zeta^N \ ,
\eeq
which generalizes the usual result for the second derivatives of Killing vectors.

Noether currents should now satisfy the \emph{modified conservation laws}
\beq
D_M\,{\cal J}^M \ - \ {S^M}_{MN}\, {\cal J}^N \ = \ 0 \ , \label{mod_conserv}
\eeq
a subtlety whose origin we already highlighted in eq.~\eqref{tot_der}.  Given a Killing vector $\zeta^A$ solving eq.~\eqref{mod_killing}, one can indeed verify that
\beq
{\cal J}^M \ = \ {{\cal T}^M}_N\, \zeta^N \ - \ {{\cal Y}^M}_{AB}\, \theta^{AB} \ , \label{noether}
\eeq
with $\theta^{AB}$ given by eq.~\eqref{epsilonAB},
satisfies the modified conservation law~\eqref{mod_conserv}. To this end, notice that
the Bianchi identities of eqs.~\eqref{bianchi1} and \eqref{bianchi2} give
\bea
D_M\,{\cal J}^M \ - \ {S^P}_{PM} \,{\cal J}^M\!\! &=& \!\! - \left({S^P}_{MN}\,{{\cal T}^M}_P \ + \ {R_{MN}}^{AB}\ {{\cal Y}^M}_{AB}\right) \zeta^N \nonumber \\ &+& \!\!\! {\cal T}^{AB}\left( D_A\,\zeta_B \ - \ \theta_{AB}\right)  \, - \,  {{\cal Y}^M}_{AB}\,\emph{} D_M\,\theta^{AB} \ ,
\eea
while using the definition of $\theta^{AB}$ this expression reduces to
\bea
D_M\,{\cal J}^M \ - \ {S^P}_{PM} \,{\cal J}^M\!\! &=& \!\! - \ {{\cal Y}^M}_{AB}\left( {R_{MN}}^{AB}\, \zeta^N \ + \  D_M\,\theta^{AB}\right) \ ,
\eea
whose right--hand side vanishes on account of the second of eqs.~\eqref{deltaeomega}. Repeating considerations made in Section~\ref{sec:bianchi_b} one can now conclude that, if the modified conservation laws~\eqref{mod_conserv} are supplemented by the boundary conditions
\beq
\left.{}^{} {{\cal J}^{\,r}} \right|_{\partial\,{\cal M}} \ = \ 0  \ , \label{bcs}
\eeq
the corresponding charges
are conserved even in the presence of a boundary $\partial{\cal M}$.

\section{\sc  Lower--Dimensional Spinors from an Interval}\label{sec:applications}
In~\cite{ms2} we shall explore families of $D$--dimensional warped metrics of the type
\beq
ds^{\,2}\ = \ e^{2B(r)}dr^2\ + \ e^{2A(r)}\, g_{\mu\nu}(x)\, dx^\mu\,dx^\nu \ + \ e^{2C(r)}\, g_{ij}(y)\, d y^i\,d y^j \ , \label{metric_sym}
\eeq
where $g_{\mu\nu}$ is typically a Minkowski metric $\eta_{\mu\nu}$ of dimension $d$ and $g_{ij}$, the metric of an internal compact space of dimension $N$, is typically $\delta_{ij}$. Examples of this type were also recently described in~\cite{ab}, and a wide portion of these solutions involve, just as the ones in~\cite{dm_vacuum}, $r$-intervals of finite length. When $g_{\mu\nu}$ and $g_{ij}$ are flat metrics, the relevant Killing symmetries are translations in spacetime and along an internal torus, together with spacetime Lorentz rotations.
The former correspond to constant $\zeta^\mu$ or $\zeta^i$, so that
\beq
{\cal J}^M \ = \ {{\cal T}^M}_\mu\, \zeta^\mu \ + \  {{\cal T}^M}_i\, \zeta^i  \ , \label{noether_trans}
\eeq
while the latter correspond to
$\zeta^\mu \ = \ \theta^{\mu\nu} x_\nu$,
with constant antisymmetric $\theta^{\mu\nu}$, so that
\beq
{\cal J}^M \ = \ {{\cal T}^M}_\nu\, \theta^{\nu\rho}\,x_\rho \ - \ {{\cal Y}^M}_{\mu\nu}\, \theta^{\mu\nu} \ . \label{noether_rot}
\eeq
For the currents in eqs.~\eqref{noether_trans} and \eqref{noether_rot}, the conditions in eq.~\eqref{bcs} therefore demand that
\beq
\left. {}^{}{{\cal T}^r}_\mu\right|_{\partial\,{\cal M}} \ = \ 0 \ , \qquad \left. {}^{}{{\cal T}^r}_i \right|_{\partial\,{\cal M}} \ = \ 0 \ , \qquad \left. {}^{}{{\cal Y}^r}_{\mu\nu} \right|_{\partial\,{\cal M}} \ = \ 0 \ . \label{cond_gen}
\eeq

For a spin--$\frac{1}{2}$ fermion, ${\cal T}$ and ${\cal Y}$ are given in eq.~\eqref{TY_onehalf}, and the first two sets of conditions are implied by eqs.~\eqref{Lambda_mat} and \eqref{Lambda_constr_kin}. The last set
 puts on $\Lambda$ the additional constraints
\beq
\left.{}^{} \bar{\lambda} \, \gamma^r\,\gamma_{\mu\nu} \, \lambda\right|_{\partial{\cal M}} \ = \ 0 \ ,
\eeq
which are also solved by a matrix $\Lambda$ in eqs.~\eqref{Lambda_mat} and \eqref{Lambda_constr_kin}, provided
\beq
\left[ \Lambda , \gamma_{\mu\nu} \right] \ = \ 0 \ .\label{constraints_Lambda}
\eeq
In settings of interest for Supergravity and String Theory, $\Lambda$ is often subject to further restrictions. If the dimension $D$ of ${\cal M}$ is even and $\lambda$ is a Weyl spinor, one should demand that
\beq
\left[ \Lambda \,,\, \gamma_\chi \right] \ = \ 0 \ , \label{weyl_constr}
\eeq
where $\gamma_\chi$ is the chirality matrix of ${\cal M}$, while if $\lambda$ is a Majorana spinor one should demand that
\beq
C^{-1} \, \Lambda^T \, C \ = \ - \ \gamma^0 \, \Lambda \, \gamma^0 \ , \label{maj_constr}
\eeq
where $C$ is the charge--conjugation matrix of ${\cal M}$. When $D$ is $odd$, with no other internal manifold, the choice $\Lambda = \gamma^r$, which rests on the chirality matrix of $\partial{\cal M}$, satisfies eqs.~\eqref{Lambda_mat}, \eqref{Lambda_constr_kin}, \eqref{maj_constr} and commutes with all spacetime Lorentz generators of $\partial{\cal M}$. This case is central to the Horava--Witten construction~\cite{HW}. When $D$ is $even$, similar settings obtain with non--chiral spinors. For example, in type--IIA supergravity the choice $\Lambda=\gamma^r$, used in~\cite{pw}, respects all Lorentz symmetries in nine dimensions while connecting the two chiralities on $\partial{\cal M}$, and the Neveu--Schwarz--Ramond open string~\cite{NSR} was a first example of this type.
The situation becomes less conventional when starting from chiral spinors, which is the case for the solutions in~\cite{dm_vacuum}. Now the choice $\Lambda=\gamma^r$ violates the Weyl constraint~\eqref{weyl_constr}, so that no solutions exist that respect the whole nine--dimensional Lorentz symmetry. However, when a compact internal manifold is also present, the Weyl constraint can be solved combining $\gamma^r$ with an \emph{odd} number of internal $\gamma$'s, and a first option also compatible with the Majorana constraint~\eqref{maj_constr}, as needed in~\cite{dm_vacuum}, is $\Lambda=\gamma^6\gamma^7\gamma^8\gamma^r$. It respects the six--dimensional Lorentz group, which suffices when ${\cal I}$ combines with a three--torus.

In general, in $D$--dimensional spacetimes of ``mostly plus'' Minkowski signature,
\begin{equation}
(i)^{{n(n-1)\over 2}}\,\gamma^{A_1,\dots A_n}\ , \qquad n=0,\dots,\tilde D \ , \label{eq56}
\end{equation}
with $\tilde D=D$ if $D$ is even or $\tilde D=\frac{(D-1)}{2}$ if $D$ is odd, are a basis for $2^{\left[{D\over 2}\right]} \times 2^{\left[{D\over 2}\right]}$ matrices.
The matrices in eq~\eqref{eq56} are self-adjoint and square to one when all $A_i\neq 0$, and otherwise they are self-adjoint and square to one when multiplied by $i$. One can distinguish the two sets
\begin{equation}
(i)^{{n(n+1)\over 2}}\gamma^{r i_1,\dots i_n} \qquad {\rm and} \qquad  i(i)^{{(m+d-1)(m+d)\over 2}}\gamma^{01\dots (d-1) i_1,\dots i_m} \ ,
\end{equation}
with $n\leq \mathrm{min}(N,\tilde D-1)$ and $m+d \leq \mathrm{min}(N+d,\tilde D)$, which we call \emph{n--type} and \emph{m--type} matrices, all of which satisfy the constraints~\eqref{Lambda_constr_kin}. When $D$ is even, one can also start from a Weyl fermion, but eq.~\eqref{weyl_constr} then demands that $n+1$  and/or $m+d$ be even. Moreover, when $D=2,3,4$ modulo 8, the Majorana constraint is possible, and eq.~\eqref{maj_constr} then demands that $n=0,3,4,7$ modulo 8 or $m+d=2,3,6,7$ modulo 8. Alternatively, when $D=2,8,9$ modulo 8 the pseudo--Majorana constraint is possible and allows the same options. Finally, when $D=2$ modulo 8 the Weyl-Majorana constraint is possible~\cite{GSO}, and eq.~\eqref{maj_constr} then demands that $n=3,7$ modulo 8 or $m+d=2,6$ modulo 8. In particular, the example given above eq.~\eqref{eq56} rests on an \emph{n--type} $\Lambda$ with $n=3$. In conclusion, when starting in $D=11$ with a Majorana spinor, there are \emph{n--type} $\Lambda$'s with $n=0, 3, 4$, and \emph{m--type} $\Lambda$'s with $m+d=2,3$, because $\tilde D=5$. Moreover, when starting in $D=10$ with a Weyl spinor, there are \emph{n--type} $\Lambda$'s with $n$ odd and \emph{m--type} $\Lambda$'s with $m+d$ even. Finally, when starting in  $D=10$ with a Majorana--Weyl spinor, there are \emph{n--type} $\Lambda$'s with $n=3,7$ and \emph{m--type} $\Lambda$'s with $m+d=2, 6$. These solutions are compatible with the Lorentz symmetry in six or fewer dimensions.

A gravitino $\psi_M$ contains lower--dimensional spin--$\frac{3}{2}$ modes $\psi_\mu$ in its space--time components, which are selected by the additional constraint
\beq
\gamma^\mu\, \psi_\mu \ = \ 0 \ ,
\eeq
to which the preceding considerations apply almost verbatim. There are also internal spin--$\frac{1}{2}$ components that mix, in general, with other spinor modes. For example, the internal component of a Majorana--Weyl gravitino in nine dimensions yields a spinor of chirality opposite to the one present in the ten--dimensional $(1,0)$ supergravity multiplet. The two build a Majorana spinor, so that at the ends of ${\cal I}$ one can relate them with $\Lambda=\gamma^r$, but the other Fermi modes of the Sugimoto model~\cite{bsb} do not satisfy the boundary conditions \eqref{Lambda_mat} compatibly with the full Lorentz symmetry of more than six non--compact dimensions.
Notice, finally, that different choices of $\Lambda$ at the two ends of ${\cal I}$ could be used, in general~\cite{dud_groj}, to induce Scherk--Schwarz deformations~\cite{ss}.

These considerations have counterparts in $AdS_{2n}$, which have a boundary at infinity, so that, in view of the preceding discussion, chiral fermions are not compatible with their isometries. The chiral limit of a massive fermion propagator is indeed singular in $AdS_4$, while the order parameter $\langle \bar{\psi} \psi\rangle$ acquires a vacuum value inversely proportional to the $AdS$ radius~\cite{all_lut}.

\vskip 18pt
\section*{Acknowledgments}
We are grateful to Sergio Ferrara for a stimulating discussion. AS was supported in part by Scuola Normale, by INFN (IS GSS-Pi) and by the MIUR-PRIN contract 2017CC72MK\_003. JM is grateful to Scuola Normale Superiore for the kind hospitality, while AS is grateful to U. Paris VII and DESY-Hamburg for the kind hospitality, and to the Alexander von Humboldt Foundation for the generous support, while this work was in progress.

\end{document}